\documentclass{article}

\usepackage{arxiv}

\usepackage[utf8]{inputenc} 
\usepackage[T1]{fontenc}    
\usepackage{hyperref}       
\usepackage{url}            
\usepackage{booktabs}       
\usepackage{amsfonts}       
\usepackage{nicefrac}       
\usepackage{microtype}      
\usepackage{lipsum}		
\usepackage{graphicx}
\usepackage{natbib}
\usepackage{doi}
\def\blambda{{\boldsymbol{\lambda}}}

\usepackage{amssymb,amsmath}
\usepackage{amsthm}
\usepackage{tabularx}
\usepackage{enumerate}
\usepackage{float}
\usepackage{bm}
\usepackage{placeins}
\usepackage{rotating}
\usepackage{sistyle}
\usepackage{morefloats}
\usepackage{setspace}
\usepackage{amsmath}
\usepackage{threeparttable}
\usepackage{sistyle}
\usepackage{graphics}
\usepackage{url}
\usepackage{amssymb}
\usepackage{booktabs}
\usepackage{multirow}
\usepackage[normalem]{ulem} 
\usepackage{tcolorbox}
\usepackage{blkarray}
\usepackage{caption}
\usepackage{subcaption}
\usepackage[export]{adjustbox}

\allowdisplaybreaks
\SIthousandsep{,}

\newcommand{\bZ}{\pmb{Z}}

\title{Correcting Conditional Mean Imputation for Censored Covariates and Improving Usability}

\author{\href{https://orcid.org/0000-0001-5380-2427}{\includegraphics[scale=0.06]{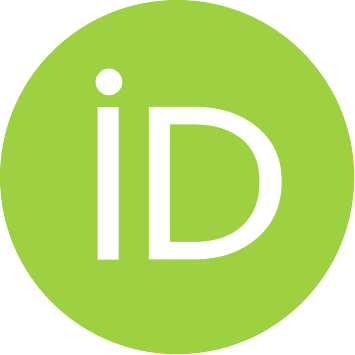}\hspace{1mm}Sarah C.~Lotspeich}\thanks{These authors contributed equally to this work.}\\
	Department of Biostatistics\\
	University of North Carolina at Chapel Hill\\
	Gillings School of Global Public Health\\
	Chapel Hill, NC, U.S.A. \\
	\And
	\href{https://orcid.org/0000-0001-5927-3968}{\includegraphics[scale=0.06]{orcid.pdf}\hspace{1mm}Kyle F. Grosser$^*$}\\
	Department of Biostatistics\\
	University of North Carolina at Chapel Hill\\
	Gillings School of Global Public Health\\
	Chapel Hill, NC, U.S.A. \\
    \And
    \href{https://orcid.org/0000-0001-9900-4025}{\includegraphics[scale=0.06]{orcid.pdf}\hspace{1mm}Tanya P.~Garcia}\\
	Department of Biostatistics\\
	University of North Carolina at Chapel Hill\\
	Gillings School of Global Public Health\\
	Chapel Hill, NC, U.S.A. \\
    \texttt{tpgarcia@email.unc.edu} \\
}

\renewcommand{\shorttitle}{Correcting Conditional Mean Imputation for Censored Covariate}

\hypersetup{
pdftitle={Correcting Conditional Mean Imputation for Censored Covariates and Improving Usability},
pdfsubject={q-stat.ME},
pdfauthor={Sarah C. ~Lotspeich, Kyle F. ~Grosser, Tanya P. ~Garcia},
pdfkeywords={Bootstrap, Limit of detection, Longitudinal data, Multiple imputation, Survival analysis},
}

\begin{document}
\maketitle

\begin{abstract}
Analysts are often confronted with censoring, wherein some variables are not observed at their true value, but rather at a value that is known to fall above or below that truth. While much attention has been given to the analysis of censored outcomes, contemporary focus has shifted to censored covariates, as well. Missing data is often overcome using multiple imputation, which leverages the entire dataset by replacing missing values with informed placeholders, and this method can be modified for censored data by also incorporating partial information from censored values. One such modification involves replacing censored covariates with their conditional means given other fully observed information, such as the censored value or additional covariates. 
So-called conditional mean imputation approaches were proposed for censored covariates in \citet{AtemEtAl2017}, \citet{AtemSampeneGreene2019}, and \citet{AtemMatsoukaZimmern2019}. These methods are robust to additional parametric assumptions on the censored covariate and utilize all available data, which is appealing. As we worked to implement these methods, however, we discovered that these three manuscripts provide nonequivalent formulas and, in fact, none is the correct formula for the conditional mean. Herein, we derive the correct form of the conditional mean and demonstrate the impact of the incorrect formulas on the imputed values and statistical inference. Under several settings considered, using an incorrect formula is seen to seriously bias parameter estimation in simple linear regression. Lastly, we provide user-friendly {\tt{R}} software, the {\tt{imputeCensoRd}} package, to enable future researchers to tackle censored covariates in their data.
\end{abstract}


\section{Challenges with Current Approaches}

In studies of longitudinal data, analysts are often confronted with censoring, wherein some variables are not observed at their true value. Instead, the \textit{censored} value, which is known to fall above or below that truth depending on the type of censoring, is observed in its place. Suppose we are interested in age at onset of cardiovascular disease; if a subject drops out of our study at age 50 without developing cardiovascular disease, then the variable has been censored since we know only that the age of onset must be greater than 50. While it might be tempting to describe this variable as missing, it would be more informative (and appropriate) to say that it was censored, which incorporates partial information that, for example, age of onset cannot be less than or equal to 50. While conventional survival analyses most often focus on censored outcomes, more contemporary analyses extend to accommodate censored covariates, as well. Continuing with our example, suppose we are interested in evaluating an offspring's risk of cardiovascular disease, measured by low-density lipoprotein cholesterol (outcome), as a function of age of parental onset of the same disease (covariate that is potentially censored) \citep{Mahmood2014}. Not every offspring's parent will develop the disease during follow-up, so this covariate will be incompletely observed (i.e., censored) for such subjects. In this note, we focus on right censoring, wherein true values are known to fall above censored ones, as in this example. 

A popular way to overcome conventional missing data while leveraging the entire dataset is imputation, whereby missing values are replaced by informed placeholders. Thus, completed (i.e., imputed) datasets are constructed on which standard complete-data methods, such as least-squares estimation for normal linear regression, can be applied. Imputation methods can similarly be applied to censored covariates. However, in this setting, partial information (i.e., the right-censored value which is known to be less than the true covariate) is available which would not be with traditional missing data. To capture this partial information, censored covariates can be replaced with their conditional means given other fully observed information such as the censored value or additional covariates; we refer to this approach as conditional mean imputation. A mean imputation which does not take into account the nature of the censoring might yield values that are not possible; for example, an imputed value lower than the right-censored one would be unrealistic. 

Recently, such a conditional mean imputation approach was proposed for censored covariates in \citet{AtemEtAl2017}, \citet{AtemSampeneGreene2019}, and \citet{AtemMatsoukaZimmern2019}. This conditional mean, as shown below, requires estimating the survival function for the censored covariate. If misspecified, as could happen with a parametric survival model, the resulting estimates can be biased and the variability inflated. To circumvent this, Atem and colleagues modeled the survival function with the nonparametric Kaplan-Meier estimator or semiparametric Cox proportional hazards model; the former is applicable only without additional fully observed covariates, while the latter incorporates these. The Kaplan-Meier estimator makes no distributional assumptions, and the Cox model captures additional covariate effects without requiring a fully-specified distribution for the censored covariate. 

That the methods of \citet{AtemEtAl2017}, \citet{AtemSampeneGreene2019}, and \citet{AtemMatsoukaZimmern2019} are robust to additional parametric assumptions and make the most of the available dataset makes them very desirable. However, in an effort to implement them, we found that these three manuscripts actually compute the conditional mean incorrectly and provide nonequivalent formulas. Therefore, to remedy this and enable others to readily implement this method, we derive the correct form of the conditional mean and provide user-friendly {\tt{R}} software, the {\tt{imputeCensoRd}} package. We also illustrate the bias created when using the incorrect formulas.

\section{What Is the Correct Form of the Conditional Mean?}

To replicate the setup in the methods of Atem and colleagues, we consider a linear regression model, ${\rm E}(Y|X, \bZ) = \alpha + \beta X + \pmb{\gamma}^T\bZ$, relating the outcome $Y$ to censored and fully observed covariates $X$ and $\bZ$, respectively. Data are available on a sample of $n$ subjects, but in place of $X$ is the observed covariate value $T = \min(X, C)$, where we assume $C$ is the noninformative, right-censored value and define the event indicator $\delta = {\rm I}(X \leq C)$. Noninformative censoring, i.e., the assumption that $X$ is independent of $C$, is necessary for $X$ to be missing at random \citep{Little2002} from censored subjects, and it dictates the form of the expectation that follows (as noted). We focus on right censoring since that is the setting of the existing works, but modifications for left censoring follow closely from the methods provided. Conditional mean imputation involves computing ${\rm E}(X|X>C,\bZ)$ for each censored subject, which we prove below is equivalent to
\begin{align}
{\rm E}(X|X>C_i,\bZ_i) &= C_i + \frac{1}{S_{x}(C_i|\bZ_i)}\int_{C_i}^{\infty}S_{x}(x|\bZ_i){\rm d}x \nonumber \\
&= C_i + \frac{1}{S_{0}(C_i)^{\exp(\blambda^T\bZ_i)})}\int_{C_i}^{\infty}S_{0}(x)^{\exp(\blambda^T\bZ_i)}{\rm d}x, \label{mean}   
\end{align}
where $S_{x}(t|\bZ) = {\rm P}(X \geq t|\bZ)$ is the conditional survival function of $X|\bZ$, $S_{0}(t) = S_{x}(t|\bZ = \pmb{0})$ is the baseline survival function, and $\blambda$ are the log hazard ratios from the Cox model for $X|\bZ$.

Under right censoring, the true covariate for a censored value $T_i$ ($i:\delta_i = 0$) is known to be greater than $T_i$. One way to compute the integral in Equation~\eqref{mean} is using the trapezoidal rule, incorporating the indicator ${\rm I}(T_{(j)} \geq C_i)$ for right censoring, which leads to the estimated version of the conditional mean as $\widehat{\rm E}(X_i|X_i>C_i,\bZ_i)$
\begin{align}
&= C_i + \frac{1}{2}\left[\frac{\sum_{j=1}^{n-1}{\rm I}(T_{(j)} \geq C_i)\left\{S_{0}\left(T_{(j+1)}\right)^{\exp(\blambda^T\bZ_i)}+ S_{0}\left(T_{(j)}\right)^{\exp(\blambda^T\bZ_i)}\right\}\left(T_{(j+1)}-T_{(j)}\right)}{S_{0}(C_i)^{\exp(\blambda^T\bZ_i)}}\right],\label{mean_approx}
\end{align}
where $T_{(1)} < T_{(2)} < \dots < T_{(n)}$ are the ordered, observed values of $T = \min(X, C)$. 

To derive Equation~\eqref{mean}, we will use rules of conditional probability, integration by parts, and the known relationship between the density and survival function of $X$. We begin simply with the definition of expectation, which gives us that ${\rm E}(X|X>C_i,\bZ_i) = \int_{-\infty}^{\infty}x{\rm P}(x|x>C_i,\bZ_i){\rm d}x$, where only $X$ is considered random given observed data ($C_i, \bZ_i$) for subject $i$ ($i = 1, \dots, n$). It follows from the rules of conditional probability that ${\rm P}(X|X>C_i,\bZ_i) = {\rm P}(X,X>C_i|\bZ_i)/{\rm P}(X>C_i|\bZ_i)$, where we recognize ${\rm P}(X>C_i|\bZ_i)$ as the conditional survival function of $X|\bZ = \bZ_i$ at time $C_i$ and thus ${\rm P}(X|X>C_i,\bZ_i) = {\rm P}(X,X>C_i|\bZ_i)/S_{x}(C_i|\bZ_i)$. This allows the denominator to now be factored out of the integral, leaving 
\begin{align}
{\rm E}(X|X>C_i,\bZ_i) &= \frac{1}{S_{x}(C_i|\bZ_i)}\int_{-\infty}^{\infty}x{\rm P}(x,x>C_i|\bZ_i){\rm d}x. \nonumber
\end{align}
Under the assumption of noninformative right censoring, whereby $X$ is independent of $C$,
\begin{align}
{\rm E}(X|X>C_i,\bZ_i) &= \frac{1}{S_{x}(C_i|\bZ_i)}\int_{C_i}^{\infty}x{\rm P}(x|\bZ_i){\rm d}x. \label{before_integral}
\end{align}
Now, we solve for the following using integration by parts:
\begin{align}
\int_{C_i}^{\infty}x{\rm P}(x|\bZ_i){\rm d}x &= \left\{-xS_{x}(x|\bZ_i)\right\}\Bigg|_{x=C_i}^{x\to\infty} + \int_{C_i}^{\infty}S(x|\bZ_i){\rm d}x \nonumber \\
&= \left\{\lim_{x \to \infty}-x S_{x}(x|\bZ_i)\right\} - \left\{-C_iS_{x}(C_i|\bZ_i)\right\} + \int_{C_i}^{\infty}S(x|\bZ_i){\rm d}x \nonumber \\
&= C_iS_{x}(C_i|\bZ_i) + \int_{C_i}^{\infty}S(x|\bZ_i){\rm d}x. \label{integral}
\end{align}
Plugging Equation~\eqref{integral} into Equation~\eqref{before_integral} yields
\begin{align}
{\rm E}(X|X>C_i,\bZ_i) &= \frac{1}{S_{x}(C_i|\bZ_i)}\left\{C_iS_{x}(C_i|\bZ_i) + \int_{C_i}^{\infty}S(x|\bZ_i){\rm d}x\right\} \nonumber \\
&= C_i + \frac{1}{S_{x}(C_i|\bZ_i)}\int_{C_i}^{\infty}S(x|\bZ_i){\rm d}x. \nonumber 
\end{align}
Since censored $X$ is modeled with a Cox model, we have that $S_{x}(t|\bZ_i) = S_{0}(t)^{\exp(\blambda^T\bZ_i)}$. With this, we arrive at Equation~\eqref{mean}.

\section{Impacts of Imputing with Incorrect Conditional Means}

We now discuss how the formulas from Atem and colleagues used for conditional mean imputation differ from ours in Equation~\eqref{mean_approx} and the resulting impact of these differences. 

\subsection{Misplacing the hazard ratio}\label{subsec:misplace_hr}

Compared to the correct imputed mean in Equation~\eqref{mean_approx}, the formula provided in \citet{AtemEtAl2017} is $\widehat{\rm E}(X|X>C_i,\bZ_i) =$
\begin{align}
& C_i + \frac{1}{2}\left[\frac{\sum_{j=1}^{n-1}\boxed{{\rm I}(T_{(j)} > C_i)}\left\{S_{0}\left(T_{(j+1)}\right)+ S_{0}\left(T_{(j)}\right)\right\}\boxed{^{\exp(\blambda^T\bZ_i)}}\left(T_{(j+1)}-T_{(j)}\right)}{S_{0}(C_i)^{\exp(\blambda^T\bZ_i)}}\right], \label{inc_form_AtemEtAl2017}
\end{align}
where the boxes highlight the  differences. The impact of using an indicator function with an exclusive inequality, ${\rm I}(T_{(j)} > C_i)$, rather than an inclusive inequality, ${\rm I}(T_{(j)} \geq C_i)$, is discussed in Section~\ref{subsec:underest_integral}. For now, we consider the use of the term $\left\{S_{0}\left(T_{(j+1)}\right)+ S_{0}\left(T_{(j)}\right)\right\}\boxed{^{\exp(\blambda^T\bZ_i)}}$ in the numerator, which incorrectly assumes that
\begin{align}
\left\{S_{0}\left(T_{(j+1)}\right)+ S_{0}\left(T_{(j)}\right)\right\}^{\exp(\blambda^T\bZ_i)} = \left\{S_{0}\left(T_{(j+1)}\right)^{\exp(\blambda^T\bZ_i)}+ S_{0}\left(T_{(j)}\right)^{\exp(\blambda^T\bZ_i)}\right\}.  \nonumber
\end{align}
This equality is only guaranteed to hold when $\blambda^T\bZ_i = \pmb{0}$, and thus Equation~\eqref{inc_form_AtemEtAl2017} will yield incorrect conditional mean imputations otherwise. By properties of the survival function, recall that $0 \leq S_{0}(t) \leq 1$ (for all $t$). With this in mind, it can be shown that 
\begin{align}
\left\{S_{0}\left(T_{(j+1)}\right)^{\exp(\blambda^T\bZ_i)}+ S_{0}\left(T_{(j)}\right)^{\exp(\blambda^T\bZ_i)}\right\} < \left\{S_{0}\left(T_{(j+1)}\right)+ S_{0}\left(T_{(j)}\right)\right\}^{\exp(\blambda^T\bZ_i)} \nonumber     
\end{align}
if $\blambda^T\bZ_i < \pmb{0}$ and
\begin{align}
\left\{S_{0}\left(T_{(j+1)}\right)^{\exp(\blambda^T\bZ_i)}+ S_{0}\left(T_{(j)}\right)^{\exp(\blambda^T\bZ_i)}\right\} > \left\{S_{0}\left(T_{(j+1)}\right)+ S_{0}\left(T_{(j)}\right)\right\}^{\exp(\blambda^T\bZ_i)} \nonumber     
\end{align}
if $\blambda^T\bZ_i > \pmb{0}$.
As such, Equation~\eqref{inc_form_AtemEtAl2017} will underestimate or overestimate the correct conditional means when $\blambda^T\bZ_i < \pmb{0}$ or $> \pmb{0}$, respectively. 
Next, the formula in \citet{AtemSampeneGreene2019} computes $\widehat{\rm E}(X|X>C_i,\bZ_i)$ as
\begin{align}
& C_i + \frac{1}{2}\left[\frac{\sum_{j=1}^{\boxed{n}}\boxed{{\rm I}(T_{(j)} > C_i)}\left\{S_{0}\left(T_{(j+1)}\right)+ S_{0}\left(T_{(j)}\right)\right\}\boxed{\exp(\blambda^T\bZ_i)}\left(T_{(j+1)}-T_{(j)}\right)}{S_{0}(C_i)\boxed{\exp(\blambda^T\bZ_i)}}\right]. \nonumber
\end{align}
Since both the numerator and denominator are being multiplied by the hazard ratio, $\exp(\blambda^T\bZ_i)$, it cancels out. Thus, their formula is actually equivalent to conditional mean imputation with $\widehat{\rm E}(X|X>C_i,\bZ_i) =$
\begin{align}
C_i + \frac{1}{2}\left[\frac{\sum_{j=1}^{\boxed{n}}\boxed{{\rm I}(T_{(j)} > C_i)}\left\{S_{0}\left(T_{(j+1)}\right)+ S_{0}\left(T_{(j)}\right)\right\}\left(T_{(j+1)}-T_{(j)}\right)}{S_{0}(C_i)}\right],\label{inc_form_AtemSampeneGreene2019}
\end{align}
effectively imputing the unobserved true covariate with ${\rm E}(X|X>C_i,\bZ_i) = C_i + \left\{\int_{C_i}^{\infty}S_{0}(x){\rm d}x\right\}/S_{0}(C_i)$. This means the conditional mean imputation following the formula from \citet{AtemSampeneGreene2019} ignores available covariate information.

The final formula comes from \citet{AtemMatsoukaZimmern2019}, which states that $\widehat{\rm E}(X|X>C_i,\bZ_i) =$
\begin{align}
C_i + \boxed{\left(\frac{1}{2}\right)^{\exp(\blambda^T\bZ_i)}}\frac{\sum_{j=1}^{\boxed{n}}\boxed{{\rm I}(T_{(j)} > C_i)}\left\{S_{0}\left(T_{(j+1)}\right)^{\exp(\blambda^T\bZ_i)}+ S_{0}\left(T_{(j)}\right)^{\exp(\blambda^T\bZ_i)}\right\}\left(T_{(j+1)}-T_{(j)}\right)}{S_{0}(C_i)^{\exp(\blambda^T\bZ_i)}}. \label{inc_form_AtemMatsoukaZimmern2019}
\end{align}
This trapezoidal approximation to the integral will be incorrect for all $\blambda^T \bZ_i \neq \pmb{0}$. Specifically, for all $\blambda^T \bZ_i < \pmb{0}$, Equation~\eqref{inc_form_AtemMatsoukaZimmern2019} will overestimate the imputed value, since $(1/2)^a > 1/2$ for all $a < 0$. Conversely, when $\blambda^T \bZ_i > \pmb{0}$ this formula systematically underestimates the imputed covariate value because $(1/2)^a < 1/2$ for all $a > 0$. In either case, the degree of over- or underestimation will become more severe as $\blambda^T\bZ_i$ deviates further from $\pmb{0}$. In addition, the upper bounds of the summands for Equations~ \eqref{inc_form_AtemSampeneGreene2019} and \eqref{inc_form_AtemMatsoukaZimmern2019} are invalid because we cannot evaluate $T_{(n+1)}$ in a sample of $n$ subjects.

\subsection{Underestimating the integral}\label{subsec:underest_integral}

We now highlight the impact of using an exclusive inequality ($>$) rather than an inclusive inequality ($\geq$) within the indicator function of the trapezoidal rule, as in Equations~ \eqref{inc_form_AtemEtAl2017}--\eqref{inc_form_AtemMatsoukaZimmern2019}. For simplicity, we assume that all observed times $T_i$ are unique. Then, the correct formula presented in Equation~\eqref{mean_approx} is equivalent to 
\begin{align*}
C_i + &\frac{1}{2}\left[\frac{\sum_{j=1}^{n-1}{\rm I}(T_{(j)} > C_i)\left\{S_{0}\left(T_{(j+1)}\right)^{\exp(\blambda^T\bZ_i)}+ S_{0}\left(T_{(j)}\right)^{\exp(\blambda^T\bZ_i)}\right\}\left(T_{(j+1)}-T_{(j)}\right)}{S_{0}(C_i)^{\exp(\blambda^T\bZ_i)}}\right] \\
+ &\frac{1}{2}\left[\frac{\left\{S_{0}\left(m_i\right)^{\exp(\blambda^T\bZ_i)}+ S_{0}\left(C_i\right)^{\exp(\blambda^T\bZ_i)}\right\}\left(m_i-C_i\right)}{S_{0}(C_i)^{\exp(\blambda^T\bZ_i)}}\right] \\
\end{align*}
where $m_i = \min\left\{T_{(j)}:T_{(j)} > C_i\right\}$, i.e., the first observed value greater than $C_i$. Thus, using the exclusive indicator ${\rm I}(T_{(j)} > C_i)$ as in \citet{AtemEtAl2017}, \citet{AtemSampeneGreene2019}, or \citet{AtemMatsoukaZimmern2019} leads to underestimation of the integral, $\int_{C_i}^{\infty}S(x|\bZ_i){\rm d}x$, and systematically biased imputation values. Systematic underestimation of this integral does not occur only if $S_{0}\left(m_i\right)^{\exp(\blambda^T\bZ_i)}= S_{0}\left(C_i\right)^{\exp(\blambda^T\bZ_i)} = 0$.

\subsection{Simulation studies}

To see how the different versions of conditional mean imputation affect bias in the imputation values and resulting inference, we conduct brief simulations. Samples of $n=1000$ subjects were created, beginning with fully observed binary covariate $Z$ which was generated from a Bernoulli distribution with ${\rm P}(Z = 1) = 0.25$. The right censoring variable, $C$, was independently generated from an exponential distribution with rate$=4$. Covariate $X$ was generated to have an exponential baseline survival function with rate$=5$ following the procedure of \citet{Bender2005}. To do so, we generated $U$ from a uniform distribution with min$=0$ and max$=1$ and constructed the baseline cumulative hazard as ${\rm H}_0(X) = -\log(U)\exp(-\lambda Z)$, from which we have $X = {\rm H}_0(X) / \lambda$. The $\lambda$ used to generate $X$ is the log hazard ratio, and settings with $\lambda = -2, -1, 0, 1, 2$ were considered, leading to an average of 55\%, 51\%, 45\%, 39\%, and 36\% censoring, respectively. Observed values were constructed as $T = \min(X, C)$. The outcome, $Y$, is calculated as $Y = 1 + X + 0.25Z + \epsilon$, with random errors $\epsilon$ generated independently from a standard normal distribution. Each setting was replicated 1000 times. 

\begin{figure}[htb]
\begin{center}
\includegraphics[width=\textwidth]{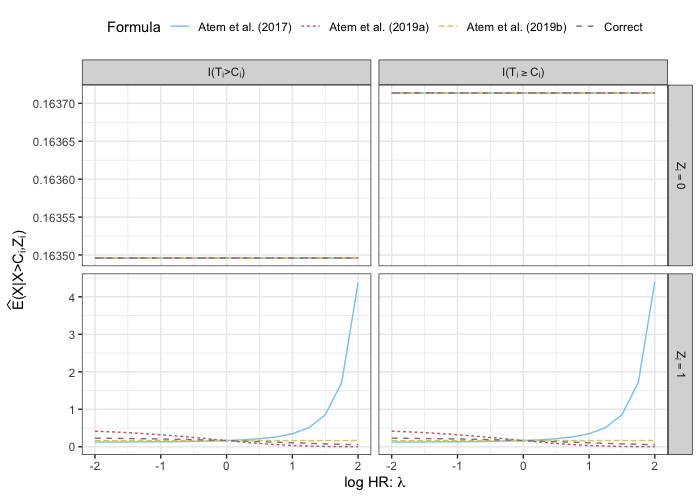}
\caption{Conditional mean imputation values, $\widehat{\rm E}(X|X>C_i,Z_i)$ for chosen censored subjects.}\label{fig:toy_example}
\end{center}
\end{figure}

Consider the conditional mean imputations from a single replication for all values of $\lambda$. We focus on the first censored subject from each stratum of binary $Z$, whose imputed values following all formulas are displayed in Figure~\ref{fig:toy_example}. Between the left and right panels, we see that the strict indicator, ${\rm I}(T_{j}>C_i)$, as in the numerator of the formulas from Atem and colleagues, led to slightly smaller conditional means due to the underestimation discussed in Section~\ref{subsec:underest_integral}. Since we know that the indicator within the trapezoidal rule approximation should use an inclusive inequality, we now focus solely on the right panel with ${\rm I}(T_{j}\geq C_i)$. With $Z = 0$, the first subject's conditional mean is a constant across the range of $\lambda$ values and all formulas agree; this is as expected since the formulas are equivalent when $\blambda^T\bZ = \pmb{0}$.

Discrepancies between the formulas are apparent for the subject with $Z = 1$. While Equation~\eqref{inc_form_AtemSampeneGreene2019} yields imputations that are still flat with respect to $\lambda$, Equation~\eqref{inc_form_AtemEtAl2017} leads to conditional means that underestimate the correct formula for $\lambda < 0$ and overestimate for $\lambda > 0$, with the degree of this overestimation seen to quickly ``blow up'' with $\lambda > 1$. Though Equation~\eqref{inc_form_AtemMatsoukaZimmern2019} is the closest to the correct imputed values, the conditional means are seen to over- and underestimate the correct imputed values when $\lambda < 0$ or $> 0$, respectively, although its deviation from the correct formula was much smaller for $\lambda > 0$. Thus, Figure~\ref{fig:toy_example} supports the implications of imputing with incorrect conditional means discussed in Sections~\ref{subsec:misplace_hr} and \ref{subsec:underest_integral}. 

Next, we investigate the ramifications of these incorrect imputed values on statistical inference about $\beta$, the coefficient for the censored covariate. We implement a multiple imputation approach wherein we iterate between (1) bootstrap resampling from the original data, (2) imputing censored covariates with conditional means, and (3) fitting a linear regression model using ordinary least-squares methods. We repeat steps (1)--(3) $B = 20$ times and ultimately pool the estimates using Rubin's rules \citep{Rubin&Schenker1991}:
\begin{align}
\hat{\beta} = \frac{1}{B}\sum_{b=1}^{B}\hat{\beta}^{(b)} \text{ and } \widehat{\rm SE}(\hat{\beta}) = \sqrt{\frac{1}{B}\sum_{b=1}^{B}\widehat{\rm Var}\left(\hat{\beta}^{(b)}\right) + \left\{ \frac{B+1}{B(B-1)}\right\}\sum_{b=1}^{B}\left(\hat{\beta}^{(b)}-\hat{\beta}\right)^2}.\nonumber
\end{align}
where $\beta^{(b)}$ is the parameter estimate obtained from the $b^{th}$ iteration and $\widehat{\rm Var}\left(\hat{\beta}^{(b)}\right)$ is the corresponding variance estimate. This multiple imputation approach follows the same steps outlined in \citet{AtemEtAl2017}, \citet{AtemSampeneGreene2019}, and \citet{AtemMatsoukaZimmern2019}. Four formulas are considered for imputation in step (2): the correct derivation (Equation~\eqref{mean_approx}) and the incorrect ones from Atem and colleagues (Equations~\eqref{inc_form_AtemEtAl2017}, \eqref{inc_form_AtemSampeneGreene2019}, and \eqref{inc_form_AtemMatsoukaZimmern2019}). In addition, we explore potential improvements due to fixing the strict indicator, i.e., replacing ${\rm I}(T_{(j)}>C_i)$ with ${\rm I}(T_{(j)}\geq C_i)$, in the incorrect formulas. 

\begin{figure}[htb]
\begin{center}
\includegraphics[width=\textwidth]{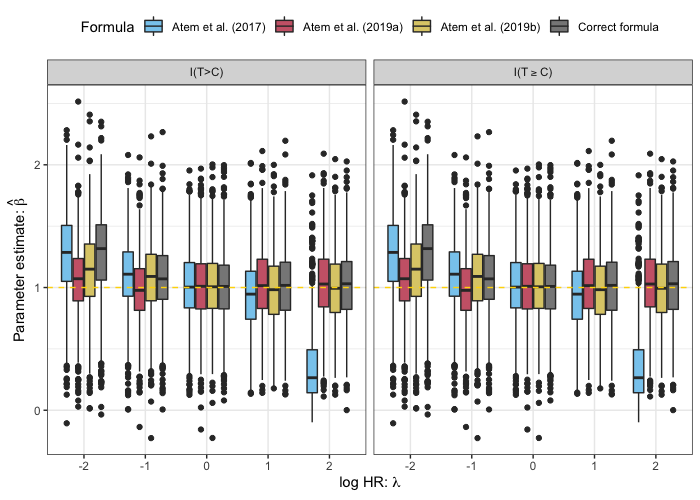}
\caption{Linear regression model estimates using conditional mean imputation values following the correct and incorrect formulas.}\label{fig:inference}
\end{center}
\end{figure}

Inference based on each of the four formulas, as well as the four formulas with ${\rm I}(T_{(j)}>C_i)$ replaced by ${\rm I}(T_{(j)}\geq C_i)$, are shown in Figure~\ref{fig:inference}. We first note, upon comparing the left and right panels, that inference does not differ greatly whether we use an exclusive indicator ${\rm I}(T_{(j)}> C_i)$ or an inclusive indicator ${\rm I}(T_{(j)}\geq C_i)$. Even though the former generates systematically smaller imputed values (Figure~\ref{fig:toy_example}), the impact on inference appears negligible. We then note the impact on inference that is caused by the choice of formula. As expected, performance was comparable when $\lambda = 0$, but differences were apparent for all other settings. Biased parameter estimates $\hat{\beta}$ were most noticeable for the formula from \citet{AtemEtAl2017}, which overestimated when $\lambda < 0$ but underestimated when $\lambda > 0$; the severity of the bias increased as $\lambda$ deviated further from zero, as well. Perhaps surprisingly, results based on the formulas from \citet{AtemSampeneGreene2019} and \citet{AtemMatsoukaZimmern2019} yielded results that were in many settings comparable to those using the correct formula. This is by no means a universal result, and the fact remains that none of the three incorrect formulas is a reliable substitute for the correct formula. These results are consistent with those under smaller sample sizes of $n = 100$ or $500$ (data not shown); there was bias for $\lambda \neq 0$ with $n = 100$ but it was seen to persist through at least $n = 1000$. 

\section{Empowering Future Usability}

In this note, we highlight between-paper disagreement between the formulas for $\widehat{\rm E}(X|X>C_i,\bZ_i)$ provided by \citet{AtemEtAl2017}, \citet{AtemMatsoukaZimmern2019}, and \citet{AtemSampeneGreene2019}
and derive its correct form. To allow others to readily use this method, we also provide software implementations for conditional mean imputation, with or without additional covariates, in the {\tt{imputeCensoRd}} {\tt{R}} package \citep{R} on our GitHub at {\tt{https://github.com/kylefred/Imputing-Censored-Covariates}}. We believe that this note and the accompanying software will help to correct and broaden adoption of the conditional mean imputation approach for covariate censoring.

\section*{Acknowledgements}
This research was supported by the National Institute of Environmental Health Sciences grant T32ES007018 and the National Institute of Neurological Disorders and Stroke (NINDS) grant K01NS099343.

\section*{Supporting Information} 
The {\tt{R}} code to replicate our simulations are available along with the {\tt{imputeCensoRd}} package on our GitHub at {\tt{https://github.com/kylefred/Imputing-Censored-Covariates}}.

\bibliographystyle{unsrtnat}
\bibliography{ref}

\end{document}